\documentclass[notoc]{JHEP3}
\usepackage{amsmath}
\usepackage{epsfig}

\title{The $Q_t$ distribution of the Breit current hemisphere in DIS
as a probe of small-$x$ broadening effects}

\author{Mrinal Dasgupta and Yazid Delenda\\
        School of Physics and Astronomy, University of Manchester,\\
        Oxford road, Manchester M13 9PL, U.K.\\
        \email{Mrinal.Dasgupta@cern.ch},~
        \email{yazid@hep.man.ac.uk}}

\preprint{MAN/HEP/2006/19}

\abstract{We study the distribution $1/\sigma\,d\sigma/d Q_t$, where
$Q_t$ is the modulus of the transverse momentum vector, obtained by
summing over all hadrons, in the current hemisphere of the DIS Breit
frame. We resum the large logarithms in the small $Q_t$ region, to
next-to--leading logarithmic accuracy, including the non-global
logarithms involved. We point out that this observable is simply
related to the Drell-Yan vector boson and predicted Higgs $Q_t$
spectra at hadron colliders. Comparing our predictions to existing
HERA data thus ought to be a valuable source of information on the
role or absence of small-$x$ (BFKL) effects, neglected in
conventional resummations of such quantities.}


\begin{document}

\section{Introduction}

With the imminent advent of the LHC, considerable effort is being
dedicated to utilising existing collider data and theoretical
predictions for QCD observables to foster an even better knowledge
of crucial parameters such as the strong coupling constant and
parton distributions in hadrons \cite{hlhc1}.

An invaluable source for such data is the HERA collider, which
continues to play an important role in phenomenological studies of
QCD. The recently concluded HERA-LHC
workshop\footnote{\href{http://www.desy.de/~heralhc}
{http://www.desy.de/~heralhc}} was in fact dedicated towards the aim
of directly linking HERA QCD studies to those that will be important
in a discovery context at the LHC.

In the present paper we shall highlight one such study. To
illustrate our point we choose the transverse momentum ($Q_t$)
spectrum of the Higgs boson for which very accurate theoretical
predictions  exist in the literature~\cite{Catani}. These estimates
resum large logarithms in the small $Q_t$ limit to next-to--next-to
leading logarithmic (NNLL) accuracy\footnote{For reference, most
resummations e.g. those for several event-shapes at LEP and HERA,
achieve only a next-to--leading (NLL) logarithmic accuracy.}. The
resummed predictions are then combined with fixed next-to--leading
order (NLO) computations~\cite{NLOHiggs} so as to achieve the best
possible accuracy over a wide range of $Q_t$. Such accurate studies
are important in the context of formulating improved strategies to
extract the signal and to enhance its statistical significance over
backgrounds.

One concern that has been a point for some discussion (see e.g.
Ref.~\cite{Nadolsky}) is the role that might be played by neglected
small-$x$ effects that one may expect could be relevant at the $x$
values involved in vector boson and Higgs production at the LHC.
Here $x$ can be interpreted, as usual, as the momentum fraction of
the incoming hadron carried by the struck parton. If neglected
small-$x$ effects are indeed important in this context, they could
in principle lead to broader $Q_t$ spectra than those predicted by
conventional resummations (sometimes also referred to as
Collins-Soper--Sterman or CSS formalism~\cite{CSS}) alone.
Henceforth we shall refer to these resummations involving $Q_t$
spectra in hadron--hadron collisions with the generic label
Drell-Yan $Q_t$ resummations.

It was observed for instance in Ref.~\cite{Nadolsky} that effects
due to small-$x$ enhancements that were suggested by
phenomenological studies~\cite{Yuan} for semi-inclusive DIS (SIDIS)
$Q_t$ distributions, could be very significant (especially in the
context of massive vector-boson production) when extrapolated to the
smaller $x$ values that will be important at the LHC. It was also
suggested in Ref.~\cite{Nadolsky} that such effects could be visible
with Tevatron (run 2) data if one concentrates on forward production
of vector bosons alone rather than integrating over all rapidities.
On the other hand, studies for many event-shape variables in the
current hemisphere of the Breit frame in DIS have been very
successful, down to moderately small $x$ values ($x \approx
10^{-2}$), based on resummation~\cite{brd,Dassalthrust,Dassaltrc}
that did not account for BFKL or formal small-$x$ effects. The
comparison to HERA data for several event-shape variables can be
found in~\cite{Dassaltrc}.

It is also clear however that event-shapes are somewhat different
from $Q_t$ spectra of the Drell-Yan type, significantly due to their
direct sensitivity to gluon emission, rather than purely through
recoil. This difference means that event-shape variables receive
non-perturbative corrections that scale as $1/Q$ \cite{DasWeb97},
where $Q$ is the DIS hard scale. These power corrections are due to
soft gluon emission alone and hence are independent of $x$. The
quantity we shall study in the present article is closer in nature
to the Drell-Yan $Q_t$ spectra since the leading non-perturbative
effects here scale as $1/Q^2$ and can be associated to what is
commonly known as ``intrinsic $k_t$'' of partons inside hadrons.
Thus one would expect any missing small-$x$ effects that appear in
the present context (perhaps as suggested in Ref.~\cite{Nadolsky} in
terms of a small-$x$ enhanced smearing of the conventional
resummation), to manifest themselves in a very similar way in the
Drell-Yan case. It is thus conceivable that for the $Q_t$
distribution we present here, deviations are seen from the resummed
form, even at $x$ values already studied successfully for
event-shape variables.

To be more precise, the observable we study here is the distribution
$1/\sigma\,d\sigma/dQ_t$, with $Q_t$ being the modulus of the
transverse momentum vector of all particles in the current
hemisphere (${\mathcal{H}_c}$) of the DIS Breit frame:
\begin{equation}
\label{Eq:defn} \vec{Q}_t = \sum_{i \in
{\mathcal{H}_c}}\vec{k}_{t,i}\, ,
\end{equation}
where $\vec{k}_t$ is measured with respect to the photon axis. Note
that the addition over particles mentioned above, is
{\emph{vectorial}} in nature, in contrast to say that involved in
event-shapes like the jet broadening observable~\cite{brd}, where
one adds the modulii of individual particle transverse momenta. This
is also a different quantity from the resummed $z$ flow studied in
Refs.~\cite{Nadolsky,Yuan} while also being directly related to the
Drell-Yan $Q_t$ spectra and thus should provide an independent probe
of small-$x$ broadening effects.

To relate the quantity here to $Q_t$ spectra in Drell-Yan type
processes one notes that in the Drell-Yan case one has the massive
vector-boson recoiling against emissions from two incoming hard
partons, while for our DIS example, the transverse momentum of the
current hemisphere is equal and opposite to that of the remnant
hemisphere. If one assumes the remnant hemisphere $Q_t$ spectrum to
be entirely generated by emissions from a single incoming leg the
relationship to Drell-Yan $Q_t$ is obvious: we just have to account
for the form-factor of one incoming parton instead of two and thus
we have, at the level of form-factors,
$\Delta^N_{\mathrm{DIS}}(Q,Q_t) =
\sqrt{\Delta^N_{\mathrm{DY}}(Q,Q_t)}$. The variable $N$ indicates
moment space, conjugate to $x$.

The simple relationship we mentioned immediately above breaks down
at the next-to--leading logarithmic (NLL) level, due to the non-global
nature~\cite{DassalNG1,DassalNG2} of the DIS observable we have
introduced. Non-globalness in the present case is a consequence of
looking at just a single hemisphere and provides an extra factor in
the DIS case, to do with soft emissions at large angles, which is
absent for the global Drell-Yan quantity. We shall of course account
for this factor, but we stress here that the fundamental
relationship of our quantity to Drell-Yan $Q_t$ (which is exactly of
the square-root form we wrote above, at leading-logarithmic
accuracy) is unchanged by this complication. In particular any
neglect of terms that are enhanced at small $x$, ought to be of
similar significance in the two cases.

To further this investigation we obtain here a resummed result for
the above quantity, to next-to--leading logarithmic (NLL) accuracy
\footnote{Equivalently we seek {\emph{single-logarithmic}} accuracy
in the resummed exponent that we shall compute subsequently.} and
combine it with fixed-order predictions to $\mathcal{O}(\alpha_s^2)$
accuracy. Our result is thus suitable for comparison with data over
the entire measured range of $Q_t$ values. We also comment on the
effect of smearing our result with a  Gaussian function, as is usual
to accommodate the so-called intrinsic $k_t$ of the incoming parton,
which has a non-perturbative origin. The resultant prediction can
then be directly compared to data which should be of interest
especially at lower $x$ values. If discrepancies are visible at low
$x$ then one may consider a small-$x$ enhanced smearing function as
was the case in Refs.~\cite{Nadolsky,Yuan}. This is certainly not a
substitute for a detailed treatment based on a physical
understanding of the small-$x$ region but merely a phenomenological
investigation into how such effects may be parameterised if present
in the first place. Subsequently one may also consider the
extrapolation of our findings to hadron colliders. We note that
preliminary data from H1 are already available~\cite{Kluge} and
await the final versions together with potential data from the ZEUS
collaboration.

The outline of this paper is as follows: in section~\ref{sec:resum}
we put together the different ingredients required to obtain our NLL
resummed result which we compute in impact parameter or $b$ space as
is most convenient for $Q_t$ resummations. Once we obtain the $b$
space answer we find, in the~\textref{sec:qt}{following section}, its
transform to $Q_t$ space and comment on its main features. In
section~\ref{sec:mat} we carry out the matching of our resummed
result to the full ${\mathcal{O}}\left(\alpha_s^2 \right )$ result
from the fixed-order program
DISENT~\cite{Nason}.~\textref{sec:np}{Finally} we comment on the
potential role of non-perturbative effects that are expected to take
the form of a smearing of the $Q_t$ distribution with a function
representing the ``intrinsic transverse momentum'' of partons inside
hadrons. Here one may try different choices for the smearing
function and search for any discrepancies at lower $x$ values,
between our results and the data. We shall leave the details of this
to our forthcoming phenomenological investigation~\cite{DasDelprep}.

\section{\label{sec:resum}Resummation}

At Born level, the struck quark is aligned along the photon axis, the
quantity in question ($Q_t$) vanishes and the distribution is
essentially a delta function: $d\sigma/dQ_t \sim \delta(Q_t)$.

At small $Q_t$ the emission of soft and collinear gluons deform the
delta function. One may, on general grounds, expect this deformation
to take the form of a Sudakov form-factor. This is essentially true
over a large range of $Q_t$ values with the caveat that at very
small $Q_t$, the correct result is no longer of Sudakov form. The
reason for this is the Parisi-Petronzio observation that the
smallest $Q_t$ values are in fact obtained by vectorial cancellation
of emissions rather than by suppressing the transverse momenta of
each individual emission \cite{PP}. We shall explain this issue more
quantitatively, with reference to our observable, in a
later~\textref{sec:qt}{section}. For now we proceed with resumming
the large logarithms that arise at small $Q_t$.

To carry out the resummation, we have to address two distinct
kinematic regimes:
\begin{itemize}
\item collinear emissions (soft or hard) along the directions of the incoming
and scattered quark directions.
\item soft emissions at large angles to the incoming and scattered (outgoing)
quark axis (since in the Breit frame the incoming and scattered
quarks are back-to--back at Born level). This contribution arises
due to the non-global nature~\cite{DassalNG1,DassalNG2} of the
observable and is a correlated multi-gluon emission term, that can
only be computed in the large $N_c$ limit.
\end{itemize}

We shall treat each region in turn starting with the collinear
enhanced contribution and then including the non-global term that
arises from the piece of the fixed-order matrix elements with only
soft enhancement (i.e. is integrable over soft gluon
{\emph{directions}}).

\subsection{Pure collinear contribution}

The collinear contribution is simple to assess since one can, to the
NLL accuracy we seek, consider collinear radiation as being included
in the evolution of the incoming and outgoing hard quark jets. In
order to derive the NLL pure collinear contribution it proves useful
to first consider the observable as defined in Eq.~\eqref{Eq:defn}.
Since one is dealing with soft and/or collinear gluon emission
alone, we are looking at a tiny deviation from the Born
configuration. Thus  the sum over current hemisphere emissions, on
the RHS of Eq.~\eqref{Eq:defn}, includes a contribution from the
transverse momentum of the outgoing current quark. To work in terms
of secondary emissions alone, one uses conservation of transverse
momentum to write Eq.~\eqref{Eq:defn} as:
\begin{equation}
\vec{Q}_t = -\sum_{i \in {\mathcal{H}_R}} \vec{k}_{t,i}\, ,
\end{equation}
where the sum now runs over all final-state particles in the
\emph{remnant} hemisphere. In the collinear region there is an
important simplification, in  that all these emissions can, to our
accuracy, be attributed to the showering of the incoming quark. Note
that since one is now inclusive over current-hemisphere emissions,
we can neglect the collinear evolution of the outgoing quark. This
will correct our resummed result by a factor of relative order
$\alpha_s$, but not enter into the NLL form-factor we aim to
compute.

The next step is to consider multiple collinear gluon branchings on
the incoming hard leg. In this region the squared matrix-element can
be approximated to NLL accuracy by a product in $N$ space of
individual gluon emissions from the hard incoming quark, where $N$
is the moment variable conjugate to Bjorken $x$. Taking first just
\emph{soft and collinear} emissions\footnote{The subsequent
extension to hard and collinear emissions will be straightforward.},
where we can just as well work in $x$ space, one can write:
\begin{equation}
\frac{1}{\sigma_0}\frac{d\sigma}{dQ_t^2} \approx \sum_n \int dP_n \,
\delta(p_t^2-Q_t^2) \, d^2 \vec{p}_t \, \delta^2 \left (
\vec{p}_t+\sum_{i \in \mathcal{H}_R}^n\vec{k}_{t,i} \right),
\end{equation}
where $dP_n$ is the differential $n$ gluon emission probability and
we introduced the vector $\vec{p}_t$, which is the vectorial sum of
transverse momenta of particles in $\mathcal{H}_R$. Note that for
purely soft and collinear emissions the parton distribution function
(pdf) cancels with that in $\sigma_0$, the Born cross-section. This
is not correct in the hard-collinear region and we will re-introduce
the pdf while considering those emissions. For the final result, we
shall also normalise to the cross-section including up to
${\mathcal{O}} (\alpha_s^2)$ corrections, rather than merely the
Born cross-section.

We first compute the integrated quantity:
\begin{equation}
\frac{1}{\sigma_0}\sigma(Q,Q_t) = \frac{1}{\sigma_0} \int_0^{Q_t^2}
\frac{d\sigma}{d Q_t'^2} dQ_t'^2 \approx \sum_n \int dP_n \, \theta
(Q_t-p_t) \,  d^2 \vec{p}_t \, \delta^2 \left ( \vec{p}_t+\sum_{i
\in \mathcal{H}_R}\vec{k}_{t,i} \right).
\end{equation}
One can then express:
\begin{equation}
\label{eq:fac1} \delta^2\left ( \vec{p}_t+\sum_{i\in
\mathcal{H}_R}{\vec{k}_{t,i}} \right ) = \int \frac{d^2
\vec{b}}{(2\pi)^2} e^{i\vec{b}.\vec{p}_t} \prod_{i\in \mathcal{H}_R}
e^{i \vec{b}. \vec{k}_{t,i}}\,.
\end{equation}
Having achieved our aim of factorising the delta function constraint
into a product of individual gluon contributions, we integrate over
$\vec{p}_t$ which reduces the above to:
\begin{equation}
\label{eq:fac2} \frac{1}{\sigma_0} \sigma(Q,Q_t) \approx \sum_n \int
dP_n\, Q_t \,J_1(b\,Q_t)\,db\prod_{i\in \mathcal{H}_R} e^{i \vec{b}.
\vec{k}_{t,i}}\, ,
\end{equation}
where in writing the above we made use of $\int_0^{2 \pi} d\theta\,
\exp\left(i\,b\,p_t \cos \theta\right)=2\pi J_0(b\,p_t)$, and $u J_1(u) =
\int_0^u u'
J_0(u') du'$ with $J_0$
and $J_1$ being the zeroth and first order Bessel functions.

The emission probability $dP_n$ factorises for \emph{soft and
collinear} emissions into an essentially classical
independent-emission pattern:
\begin{equation}
\label{eq:fac3} dP_n = \frac{1}{n!} \prod_{i\in \mathcal{H}_R}
C_F\frac{\alpha_s(k_{t,i}^2)}{\pi} \frac{d^2 \vec{k}_{t,i}}{\pi
k_{t,i}^2} d\eta_i\, ,
\end{equation}
where $\vec{k}_{t,i}$ and $\eta_i$ refer to the transverse momenta
and rapidity of the $i^{\textrm{th}}$ emission with respect to the
incoming quark direction . Since the incoming quark is anti-parallel
to the photon axis, in the Breit frame, the $k_t$ immediately above
is the same to NLL accuracy as that measured with respect to the
photon axis and we do not distinguish the two. The coupling
$\alpha_s$ is defined in the CMW scheme \cite{CMW}. We note that the
rapidity integration is bounded by $\eta=0$ at large angles to the
incoming quark since we are considering emissions in the remnant
hemisphere alone.

Summing over all emissions in the remnant hemisphere using the
factorised forms \eqref{eq:fac1} and \eqref{eq:fac3} and inserting
virtual corrections according to the emission pattern
\eqref{eq:fac3} (with an additional factor $(-1)^n$) we arrive at
the resummed soft and collinear contribution to
$\sigma(Q,Q_t)/\sigma_0$:
\begin{equation}
\label{eq:sc} \int Q_t \, J_1(b \,Q_t) \, \exp[-R_{\mathrm{sc}}(b)]
\,db\, ,
\end{equation}
where $R_{\mathrm{sc}}(b)$ is the ``radiator''  accounting for
soft and collinear emissions by the incoming quark:
\begin{equation}
\label{eq:rad} R_{\mathrm{sc}}(b) = - \frac{C_F}{\pi} \int \frac{d^2
\vec{k}_t}{\pi k_t^2} \, d\eta \, \alpha_s(k_t^2)\, \left (e^{i
\vec{b}.\vec{k}_t}-1 \right).
\end{equation}
We remind the reader that one needs to correct the above expression
to obtain single-logarithms arising from  hard collinear radiation
as well as those that arise in the large-angle region from soft
emissions, which we shall do presently.

Let us for the moment concentrate on the quantity
$R_{\mathrm{sc}}(b)$, which is the most important piece of the
result since it contains the leading (double) logarithms.
Integrating over the polar angle variable in Eq.~\eqref{eq:rad} we
obtain:
\begin{equation}
\label{eq:rad2} R_{\mathrm{sc}}(b) =- \frac{2 C_F}{\pi} \int
\frac{dk_t}{k_t} d\eta \, \alpha_s(k_t^2)\left (J_0(b\,k_t)-1
\right).
\end{equation}
Further to next-to--leading or single logarithmic accuracy, it
suffices to make the substitution  $\left (J_0(b\,k_t)-1 \right) \to
-\theta \left (k_t-2\, e^{-\gamma_E} /b\right)$ and arrive at:
\begin{equation}
\label{eq:rad3} R_{\mathrm{sc}}(b)=\frac{2 C_F}{\pi}
\int_{1/\bar{b}}^{Q} \alpha_s(k_t^2) \frac{dk_t}{k_t} \ln
\frac{Q}{k_t}\, , \qquad \bar{b} = b\,e^{\gamma_E}/2\, ,
\end{equation}
where we performed the rapidity integration.

Next we extend the soft-collinear result above to the full collinear
one by including hard emissions. As is easy to show (see
e.g.~\cite{Dassalthrust}), hard emissions collinear to the incoming
quark lead to a modification of the factorisation scale $\mu^2$ in
the pdfs to the scale $1/\bar{b}^2$, $q(x,\mu^2) \to
q(x,1/\bar{b}^2)$. A remnant of this change of scale is the
replacement of $Q$ in Eq.~\eqref{eq:rad3} by $Q\,e^{-3/4}$. Thus the
extension of the soft-collinear result, Eq.~\eqref{eq:sc}, to the
full collinear one is:
\begin{equation}
\label{eq:c} \frac{1}{q(x,\mu^2)} \int q\left(x,1/\bar{b}^2
\right)Q_t \, J_1(b\,Q_t) \, \exp[-R(\bar{b}\,Q)] db\, ,
\end{equation}
with:
\begin{equation}
\label{eq:rad10} R(\bar{b}\,Q)=\frac{2 C_F}{\pi}
\int_{1/\bar{b}}^{Q} \alpha_s(k_t^2) \frac{dk_t}{k_t} \left(\ln
\frac{Q}{k_t}-\frac{3}{4}\right).
\end{equation}

\subsection{Non-global corrections}

We also have to include the effects of soft emissions at large
angles. Thus far we have identified remnant emissions as those
belonging to the incoming quark while current hemisphere emissions
(over which we claimed to be inclusive) are associated to the struck
final-state quark. As is the case for single-hemisphere observables,
this identification is not correct at single-log level due to
correlations between soft emissions at large angles
\cite{DassalNG1,DassalNG2}. Thus the remnant hemisphere distribution
is affected at SL level by soft gluons at large angles to the
current quark, but still in the current hemisphere, emitting into
the remnant hemisphere.

The computation of this piece has been carried out, in the large
$N_c$ approximation\footnote{Strictly speaking, the full result has
been computed at $\mathcal{O}(\alpha_s^2)$ and the large $N_c$
approximation starts $\mathcal{O}(\alpha_s^3)$.} and is universal
for all observables having a linear dependence on $k_t$ of soft
large-angle emissions. We label this ``non-global'' piece
$\mathcal{S}(\bar{b}\,Q)$, which can be parameterised as
\cite{DassalNG1}:
\begin{equation}
\mathcal{S}\left(\bar{b}\,Q\right)\simeq
\exp\left\{-C_FC_A\frac{\pi^2}{3}\left
(\frac{1+(At)^2}{1+(Bt)^C}\right)t^2\right\},
\end{equation}
where:
\begin{equation}
t(\bar{b}\,Q)=-\frac{1}{4\pi\beta_0}
\ln\left(1-2\alpha_s(Q^2)\beta_0\ln(\bar{b}\,Q)\right),
\end{equation}
with $A=0.85\,C_A$, $B=0.86\,C_A$ and $C=1.33$.

Thus our final form for the resummed result to NLL accuracy can be
expressed as:
\begin{equation}
\label{eq:resb} \frac{1}{\sigma_0} \sigma(Q,Q_t) =
\frac{1}{q(x,Q^2)} \int q\left(x,1/\bar{b}^2\right )
\mathcal{S}(\bar{b}\,Q)\,e^{-R(\bar{b}\,Q)} \, Q_t \, J_1(b\,Q_t)
\,db\,,
\end{equation}
where we chose $\mu=Q$. The above result now incorporates all the
sources of logarithmic enhancements to NLL or single-log accuracy,
specifically soft and hard collinear emissions and soft emissions at
large angles. The result for the NLL radiator $R(\bar{b}\,Q)$ is
explicitly given in appendix~\ref{sec:rad}. In the following
section, we shall take the $b$ space result above and convert it to
one valid in $Q_t$ space, over the range of $Q_t$ values that
interest us.

\section{Result in $Q_t$ space}
\label{sec:qt}

We start by noting that one commonly used method to derive a $Q_t$
space result from the $b$ space form is simply to evaluate  the
complete $b$ integral in Eq.~\eqref{eq:resb} numerically.
This method is not without
several well-documented shortcomings \cite{EV,FNR} that are usually
circumvented by ``reasonable'' prescriptions that are not derived
from first principles of QCD.

For instance the $b$ integral stretches from 0 to $\infty$ but the
function $R(\bar{b}\,Q)$ has a Landau pole singularity at
$\bar{b}\,Q=\exp\left\{1/\left(2\,\beta_0\,\alpha_s\right)\right\}$,
which means it is perturbatively undefined for large $b$ values. To
get around this problem one introduces a parameter $b^{*}$ and
substitutes $b \to b^{*}=b/\sqrt{1+b^2/b_{\mathrm{lim}}^2}$
\cite{CSS}. This ensures one never evaluates $R(\bar{b}\,Q)$ or the
structure functions at scales larger than some cut-off
$b_{\mathrm{lim}}$, whose value is adjustable. Additionally, we
smear the $b$ space result with a non-perturbative Gaussian function
that is also not obtained from first principles, but typically
through fits to data sets
\cite{NPrefs1,NPrefs2,NPLadinsky,NPrefs4,NPrefs5}. These
prescriptions are needed in order to do the $b$ integral and obtain
a result for finite $Q_t$, even at relatively large $Q_t$ values
where one might expect to trust perturbative predictions and where
additional non-perturbative parameters or ad-hoc inputs should not
play any significant role. Moreover the matching to fixed-order
$Q_t$ space results is complicated by not having an analytical
resummed result in $Q_t$ space.

In what follows we provide an analytical $Q_t$ space result which is
valid for use over a large range of $Q_t$ values and represents a
clean extraction of the next-to--leading logarithmic resummed $Q_t$
space result, from the $b$ integral. The price we pay for not
evaluating the complete $b$ integral in detail, is a formal
divergence at small $Q_t$, which one can anticipate quite generally
through considerations based on the work of Parisi and Petronzio
\cite{PP}. Thus we cannot use our result at very small $Q_t$ values
since in this region our answer is no longer a good approximation to
the $b$ integral. For quantitative studies however, with the
specified $Q_t$ range over which data is available, our formula is
valid for use as it stands. The region over which we start to see a
problem with our approximation occurs at $Q_t$ values that are too
small to study accurately via perturbative methods and in any case
below the lowest $Q_t$ data.

To obtain a resummed result in $Q_t$ space we expand the function
$R(\bar{b}\,Q)$ in Eq.~\eqref{eq:resb}, about the point $\hat{b}(\equiv b\,Q_t)=2
e^{-\gamma_E}$ to obtain:
\begin{equation}
R(\bar{b}\,Q) = R \left (Q/Q_t \right) + R'(Q/Q_t) \left
(\gamma_E-\ln 2 +\ln \hat{b} \right )+\cdots\,,
\end{equation}
where we have used $R'(\bar{b}\,Q) = \partial R(\bar{b}\,Q)/\partial
\ln b$ and neglected $R''$ and higher derivatives as they contribute
only to subleading (below single-log) accuracy. The non-global
function $\mathcal{S}$ and the $b$ dependent parton distributions
are straightforwardly expressed in $Q_t$ space with the substitution
$\bar{b} \to 1/Q_t$, since logarithmic derivatives of these
single-log functions, analogous to $R'$ above, only contribute at
subleading accuracy.

Using the Taylor expansion for $R$ above, one can cast the result
\eqref{eq:resb} as:
\begin{equation}
\label{eq:res}  \frac{1}{\sigma_0} \sigma(Q,Q_t) = \frac{q\left
(x,Q_t^2\right )}{q(x,Q^2)}\mathcal{S}(Q/Q_t)\,e^{-R \left(Q/Q_t
\right)+(\ln 2-\gamma_E) R'\left (Q/Q_t \right ) }
\int_0^{\infty}d\hat{b}J_1 (\hat{b})\,\hat{b}^{-R'(Q/Q_t)}\,.
\end{equation}
Integrating over the dimensionless quantity $\hat{b}$ and
incorporating, as a factor, the $\mathcal{O}(\alpha_s)$ constant
pieces (see appendix~\ref{sec:LO}) we can write the result as:
\begin{multline}
\label{eq:res1}  \frac{1}{\sigma_0} \sigma(Q,Q_t)
=\frac{1}{q(x,Q^2)}
\left(\textbf{\emph{C}}_{0}\otimes\textbf{\emph{q}}(x,Q_t^2)
+\frac{\alpha_s}{2\pi}\textbf{\emph{C}}_{1}
\otimes\textbf{\emph{q}}(x,Q^2)\right)\times\\
\times \mathcal{S}(Q/Q_t)\,e^{-R \left(Q/Q_t \right)
 -\gamma_E R'\left (Q/Q_t \right )} \frac{\Gamma \left
 (1-R'/2 \right )}{\Gamma \left (1+R'/2 \right)}\,,
\end{multline}
where $\textbf{\emph{C}}_{0}$ and $\textbf{\emph{C}}_{1}$ are
matrices in flavour space of  coefficient functions  (see appendix
\ref{sec:LO}). While the $\textbf{\emph{C}}_{0}$ terms are merely
proportional to delta functions the $\textbf{\emph{C}}_{1}$ pieces
are important to correct the soft-collinear resummed result for hard
real and virtual emissions at the leading $\mathcal{O}(\alpha_s)$
accuracy. They are straightforward to compute and are presented in
appendix \ref{sec:LO}.

We immediately note that the above result diverges at $R'=2$, an
entirely expected feature. The reason for this divergence
(encountered also in the Drell-Yan $Q_t$ resummations and the jet
broadening in DIS \cite{brd}) is merely the fact that at very small
$Q_t$ the result one obtains is not described by exponentiation of
the leading-order result, which is essentially the form we have
derived above \cite{PP}. As $Q_t \to 0$ the mechanism of vectorial
cancellation between emissions of formally arbitrary hardness, takes
over from the Sudakov suppression of soft and collinear radiation,
as the dominant mechanism for producing a small $Q_t$. However the
divergence does not play a major role for phenomenological purposes,
since over the values of $Q_t$ we intend to study we are
sufficiently away from the point $R'=2$. This will be further
elaborated in the subsection below.

\subsection{Position and impact of the divergence}
As we mentioned above, for the particular case at hand, the
divergence occurs at rather small values of $Q_t$ for the $Q$
values\footnote{There are data in the range $17 \, \mathrm{GeV} \leq
Q \leq 116 \, \mathrm{GeV}$.} of interest to us. The corresponding
$Q_t$ values, at and near the divergence, fall in a region that is
either neglected for phenomenological purposes or modeled with the
introduction of non-perturbative parameters, since one expects
non-perturbative effects to be large here. In the $Q_t$ region where
the divergence does not have any significant numerical impact, we
still probe small enough $Q_t$ to test the perturbative resummation
and non-perturbative corrections, as is our aim.

To be precise, the divergence occurs at $R'=2$. In terms of the
variable $Q_t$, using the expression for $R'$ given in appendix
\ref{sec:rad}, this results in:
\begin{equation}
Q_t = Q \exp \left (-\frac{\pi}{\alpha_s(C_F+2 \pi \beta_0)} \right
),
\end{equation}
which for the illustrative value of  $Q = 90$ GeV gives $Q_t =0.52$
GeV, with $\alpha_s = 0.118$. Since this region is in any case
perturbatively unsafe, being quite close to the  QCD scale, we do not
expect to obtain sensible results with the perturbative methods we
have used. However we can safely study $Q_t$ values of a few GeV
without worrying about the impact of the formal divergence at
$R'=2$.

We can quantify this statement as follows: in the region where $R'
=2$, there is a complete breakdown of the hierarchy between leading,
next-to--leading, etc. logarithms. In order to determine up to which
value of $R'$ one can use the usual hierarchy, where N$^n$LL terms
are suppressed by $\alpha_s^n$ with respect to LL terms, one can
follow the procedure outlined in \cite{brd}.

From those arguments one can infer that terms that are formally NNLL
contribute a correction that is of the same order as the terms that
one keeps in the NLL resummed result, in the region where $2-R'
=\sqrt{\alpha_s}$. The NNLL terms contribute at relative
$\mathcal{O}(\alpha_s)$ when $R' =1$ or more. Thus for $R' \leq 1$
we can safely use our resummed $Q_t$ space formula since omitted
NNLL terms contribute as usual, at relative $\mathcal{O}(\alpha_s)$.

We can work out the position of both points in terms of $Q_t$ for a
given $Q$. For $Q =90$ GeV we obtain that the critical value, where
all terms in the formal hierarchy are in fact of the same order, is
$Q_t = 0.68$ GeV and that the region where the usual hierarchy is
respected, is $Q_t \geq 1.5$ GeV. This still allows the full range
of available data to be safely probed, including the lowest measured
$Q_t$ bins.

\section{Matching to fixed-order}
\label{sec:mat} Having obtained the NLL perturbative estimate we now
need to combine it with the exact ${\mathcal{O}}(\alpha_s^2)$
perturbative result to obtain accurate predictions over the entire
range where data exist. We follow the matching prescription known as
$M_2$ matching introduced in Ref.~\cite{brd}. Here, the final result
is given by:
\begin{equation}
\label{eq:matching}\sigma_r + \bar{\alpha}_s \left ( \sigma_e^{(1)}
- \sigma_r^{(1)} \right ) + \bar{\alpha}_s^2 \left
(\sigma_e^{(2)}-\sigma_r^{(2)} \right) \Sigma(Q,Q_t)\,,
\end{equation}
where $\bar{\alpha}_s =\alpha_s/(2\pi)$ and $\sigma_r^{(1),(2)}$
denote the coefficients of the resummed result, $\sigma_r$, expanded
to first and second order in $\bar{\alpha}_s$ respectively, while
$\sigma_e^{(1),(2)}$ are the corresponding coefficients obtained
from fixed-order Monte Carlo programs such as DISENT \cite{Nason}.
The above matching formula adds the resummed and exact results and
subtracts the double-counted terms (up to
$\mathcal{O}(\bar{\alpha}_s^2)$) that are included in the
resummation.

Note that terms such as $\alpha_s^2 \ln (Q/Q_t)$ that are formally
subleading and hence not included in the resummation, are present in
the piece $\bar{\alpha}_s^2 \left (\sigma_e^{(2)}-\sigma_r^{(2)}
\right)$ of Eq.~\eqref{eq:matching}. This piece is divergent as $Q_t
\to 0$ and thus we multiply it by the resummed form-factor $\Sigma(Q
,Q_t) \equiv q(x,Q_t^2)/q(x,Q^2)\,\mathcal{S}(Q/Q_t) \exp\left\{-R
\left(Q/Q_t \right)\right\}$, to ensure sensible behaviour at small
$Q_t$. This procedure is obviously ad-hoc but only affects the
result at subleading logarithmic accuracy, which is in any case
beyond our quantitative control.

Another point that needs to be re-emphasised is that the factor
$\Sigma$ as we use it here, is just an approximation to the resummed
result given by a full evaluation of the $b$ integral. The
approximation is intended for use (and is valid to NLL accuracy)
only sufficiently away from $R'=2$, the position of the divergence.
As we explained in the \textref{sec:qt}{last} section, this covers
the range over which data exist and over which we can make
reasonable comparisons.

\section{Results}
\label{sec:np}

\FIGURE[ht]{ \epsfig{file=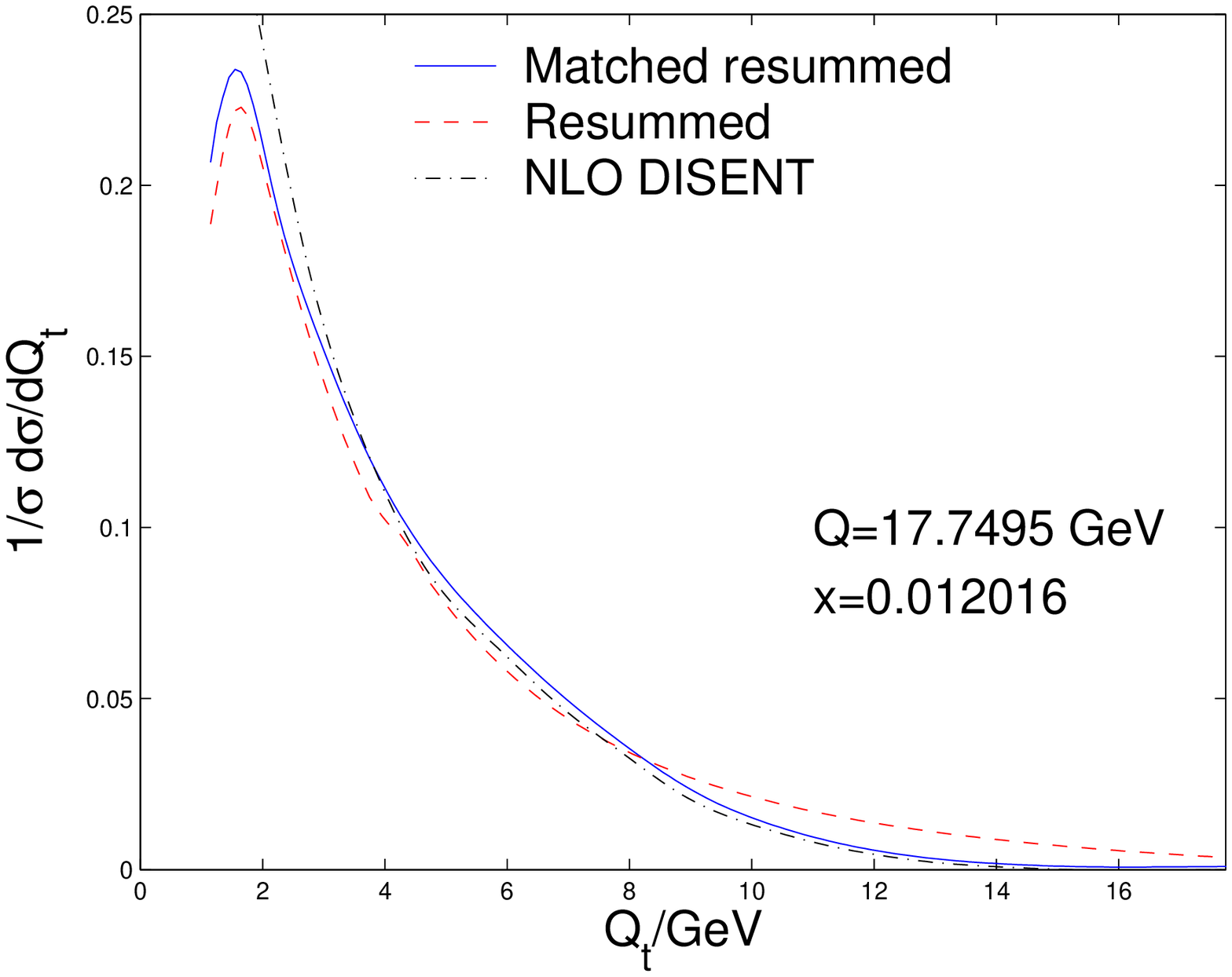,width=0.49\textwidth}
\epsfig{file=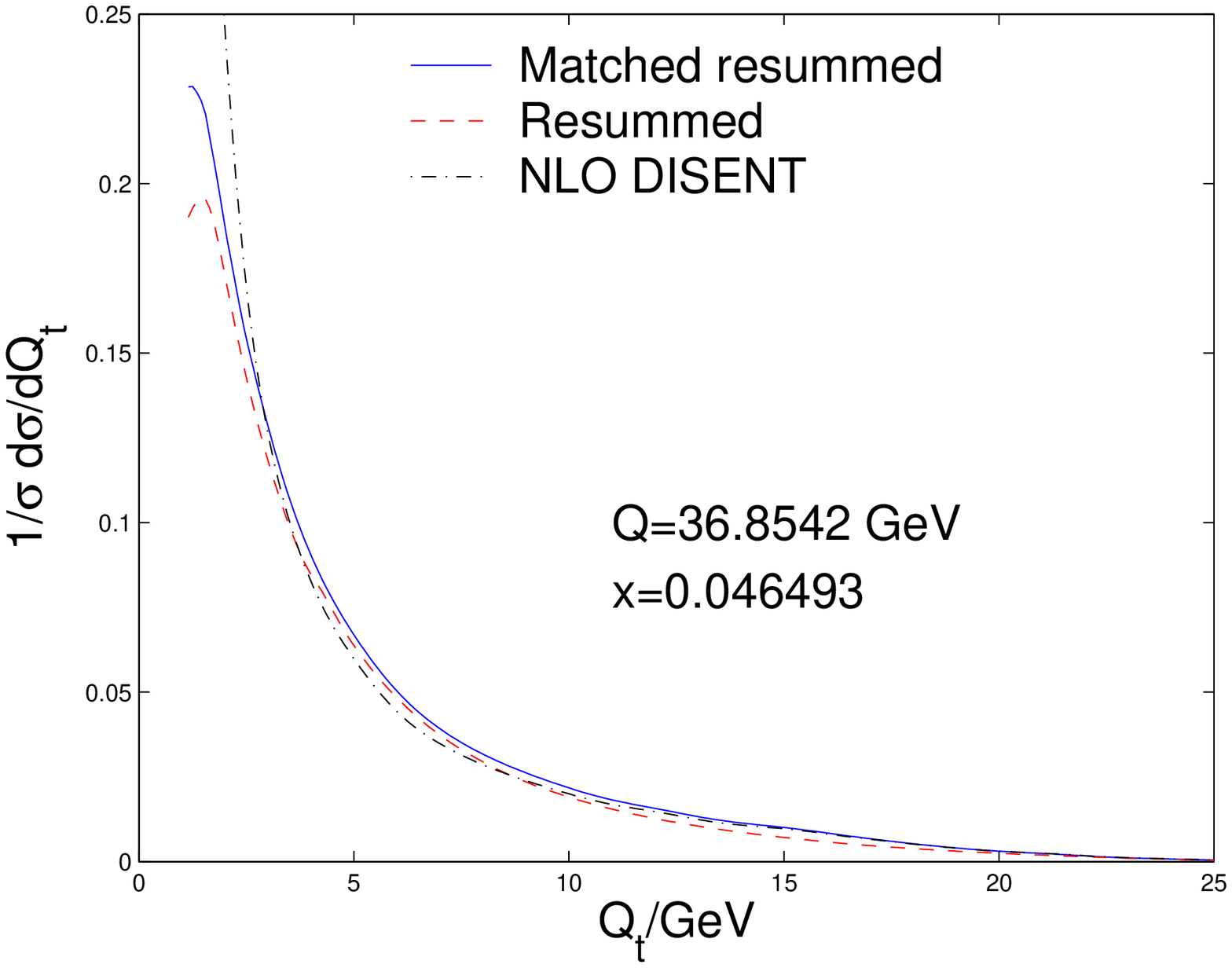,width=0.49\textwidth}
\caption{\label{fig:results}Comparison between DISENT~\cite{Nason},
matched resummed and pure resummed differential distributions. MRST
NLO pdfs have been used with $\alpha_s(M_Z^2)=0.1205$~\cite{MRST}.}}

The aim of this section is to display the results obtained for a
next-to--leading logarithmic resummed prediction matched to
next-to--leading order predictions from DISENT~\cite{Nason}.
Additionally we comment on the role that might be played by a
non-perturbative Gaussian smearing function and that small-$x$
effects may effectively give an enhanced smearing of the spectrum
leading to a broader prediction than the one provided here as was
observed also in the SIDIS case~\cite{Nadolsky,Yuan}.

In Fig.~\ref{fig:results}, we display the fixed order NLO results
along with the pure resummed and matched results for two values of
the hard scale $Q$ and corresponding Bjorken $x$ values. The choice
of the $x$ and $Q$ values correspond to bins where preliminary data
already exist~\cite{Kluge}. There one notes the divergent NLO result
and the correct small $Q_t$ behaviour as given by the resummation as
well as the role of matching in the high $Q_t$ tail of the
distribution. We also point out that the role of the non-global term
is limited to a few percent effect after the matching to fixed order
has been performed. Thus missing uncalculated single-logarithmic
terms in the non-global piece ${\mathcal{S}}$ that are suppressed as
$1/N_c^2$, are not expected to change our quantitative conclusions.
\FIGURE[ht]{ \epsfig{file=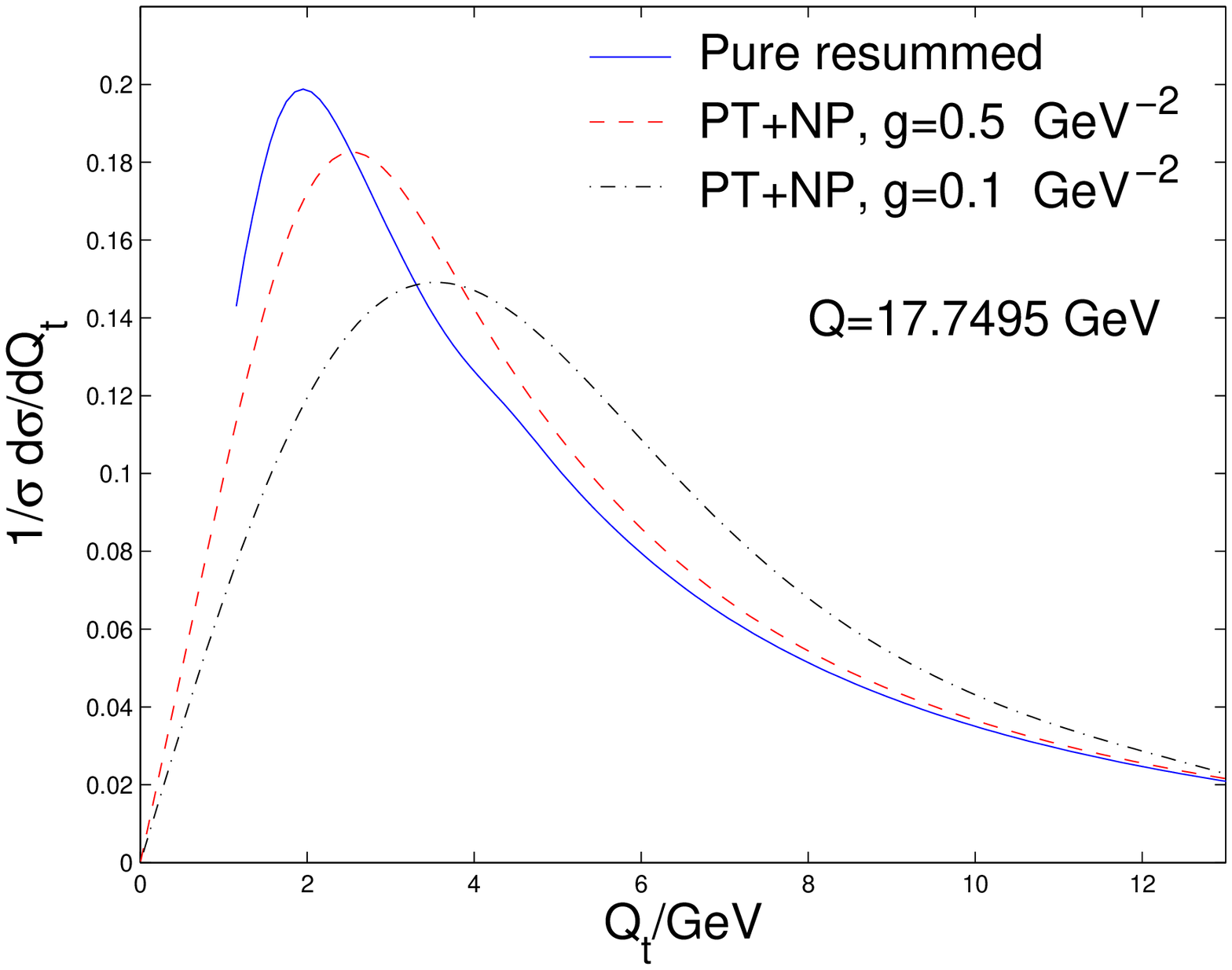,width=0.57\textwidth}
\caption{\label{fig:smear}Non-perturbative smearing of the resummed
prediction with a Gaussian function in $k_t$, $\exp[-g k_t^2]$. Two
different choices of the smearing parameter $g$ are illustrated.} }

We also present here the effect of smearing or convoluting our
resummation with a $b$ space or equivalent $k_t$ space Gaussian
function representing non-perturbative effects. Such intrinsic $k_t$
effects have been the subject of much phenomenological investigation
in Drell-Yan like processes \cite{CSS,NPLadinsky,NPSof,NPLandry}.
The non-perturbative smearing function we choose has the simple form
$F^{\mathrm{NP}}(k_t) = \exp[-g k_t^2 ] $ and we perform a
two-dimensional convolution, $\int d^2{\vec{k}_t}\,g/\pi
F^{\mathrm{NP}}\left (|\vec{Q_t}-\vec{k_t}|\right) I(k_t)$ with the
resummed distribution $I(k_t) = 1/\sigma\, d \sigma/dk_t^2$, to
obtain the smeared result as a function of $Q_t$. We carry out the
smearing with different values of $g$, two of which are illustrated
in Fig.~\ref{fig:smear}. The value of $g = 0.5\,{\mathrm{GeV}}^{-2}$
corresponds to a reasonable mean intrinsic $k_t$ value of $1.25 \,
\mathrm{GeV}$. We also illustrate the effect of using a smaller $g$
value of $0.1 \, {\mathrm{GeV}^{-2}}$, which leads as shown to a
broader $Q_t$ distribution. Ideally one would need to compare our
predictions with the data at different $x$ values in order to
ascertain whether one sees any visible broadening of the NLL
resummed spectra at smaller $x$, such as that mimicked by a change
in the smearing parameter $g$. If this is the case one may
investigate the dependence of the smearing parameter $g$ on Bjorken
$x$, a step we leave to forthcoming work \cite{DasDelprep}.

We should point out that in convoluting the resummed prediction with
the non-perturbative Gaussian $k_t$ distribution it was necessary to
provide an extrapolation of our resummed results for $I(k_t)$ down
to $k_t =0$. We have chosen this extrapolation as in Ref.~\cite{EV}
so as not to modify the NLL resummed result as well as to obtain the
correct limiting behaviour of the integrated cross-section $\propto
k_t^2, $ as determined by evaluating the $b$ integral for the
integrated quantity $\sigma(k_t)$, in the limit $k_t \to 0$.

Thus we substitute for $k_t$ an effective variable $(k_t^*)^2 =
k_t^2 + Q_{0}^2\,\exp{\left(-k_t^2/Q_0^2\right)}$ and make the
replacement $\sigma(k_t) \to \sigma(k_t^*)\left (1-\exp\left(-a
k_t^2 \right )\right )$, where $Q_0$ and $a$ should be chosen so as
not to modify the resummed result in the range where it is valid, as
in Ref.~\cite{EV}. These additional parameters should only modify
the very low (non-perturbative) $Q_t$ region where our NLL result is
in any case not valid as it stands. One can think of the parameters
$a$ and $Q_0$ as being non-perturbative inputs that can be varied
alongside the smearing parameter $g$ to obtain a good fit to data in
the lowest $Q_t$ region.

For the plot in Fig.~\ref{fig:smear} we chose to take $Q_0 = 1.2 \,
\mathrm{GeV}$ and $a =1/Q_0^2$ as these choices do not impact our
resummed results over most of their range of validity. The dominant
impact in the very low $Q_t$ region, beyond the control of NLL
resummation, is in fact that of the smearing function $\exp{(-g
k_t^2)}$, for which different choices have been already mentioned.

\section{Conclusions}
We have introduced here a DIS variable that, as we have explained,
has a very simple relationship to vector-boson and Higgs $Q_t$
spectra at hadron colliders. The aim of doing this has been to use
HERA data to compare resummed theoretical predictions with
experiment, keeping an eye on comparions at lower $x$ values. This
complements the extensive studies of DIS event shapes that have been
carried out thus far which employed the standard resummation
formalism (and non-global logarithms), supplemented by $1/Q$ power
corrections \cite{DasWeb97}. We recall that the program of comparing
DIS event shapes to the data was quite successful without any
special role visible for small-$x$ effects \cite{Dassaltrc}.

Given however that only moderately small $x$ values, $x \sim
10^{-2}$, can be reasonably accessed in these studies, it is clearly
better to choose a variable that is potentially more sensitive to
small-$x$ dynamics than event-shape variables, in order to determine
the role of these effects. We expect such a variable to be the $Q_t$
spectrum we have defined here, where it will be interesting to
establish if a small-$x$ enhanced broadening of the resummation we
presented, is indeed visible in the data. This was apparently the
case in SIDIS studies \cite{Yuan} and, if present, we expect these
effects to manifest themselves for our observable too. Given the
simple relationship of our results to those for Drell-Yan like
observables it should then be easy to extrapolate our conclusions to
hadron collider studies, where it is important to reach a firm
conclusion on the issue of small-$x$ broadening.

At present we have preliminary data for our observable from HERA and
we await the data in its final form before making detailed
comparisons and drawing phenomenological conclusions on this issue.
This will be the subject of forthcoming work.

\acknowledgments{We thank Thomas Kluge and Gavin Salam for helpful
discussions.}

\appendix

\section{Leading order result}
\label{sec:LO}

We report briefly below the leading-order result for
$\sigma_r(Q_t,Q)$, where $r$ denotes the resummed result,
neglecting terms that vanish as $Q_t \to 0$.
In line with the procedure of Ref.~\cite{brd}, the leading-order
cross-section for events with $\left|\sum_{i \in {\mathcal{H}_c}}
\vec{k}_{t,i}\right|<Q_t$ normalised to the Born cross-section
is\footnote{We compute both the $F_2$ and $F_L$ contributions
indicated by the subscripts $2$ and $L$}:
\begin{multline}\label{eq:LO}
\sigma_{r}^{(1)}(Q_t,Q,x)/\sigma_0=\frac{1}{q(x,Q^2)}
\frac{\alpha_s}{2\pi}\int_{x}^{1} \frac{d\xi}{\xi}
\Bigg\{C_F\,q\left(\frac{x}{\xi},Q^2\right)
\mathcal{F}_{q,2}\left(\xi,\frac{Q}{Q_t}\right)+\\+
T_f\,g\left(\frac{x}{\xi},Q^2\right)
\mathcal{F}_{g,2}\left(\xi,\frac{Q}{Q_t}\right)-
\frac{y^2}{1+(1-y)^2}\bigg[
C_F\,q\left(\frac{x}{\xi},Q^2\right)\mathcal{F}_{q,L}
\left(\xi,\frac{Q}{Q_t}\right)+\\
+T_f\,g\left(\frac{x}{\xi},Q^2\right)\mathcal{F}_{g,L}
\left(\xi,\frac{Q}{Q_t}\right)\bigg]\Bigg\}\,,
\end{multline}
 where $y$ is the usual DIS Bjorken variable,
$T_f=T_R\sum_{q\bar{q}}e_q^2$ and in the $\overline{\textrm{MS}}$
scheme we have:
\begin{multline}
\label{eq:fq} \mathcal{F}_{q,2}
\left(\xi,\frac{Q}{Q_t}\right)=\delta(1-\xi)
\left[-2\ln^2\frac{Q}{Q_t} +3\ln\frac{Q}{Q_t} +\frac{3}{2}\ln
2-\frac{\pi^2}{2}-\frac{9}{2} \right]
-2\left[\frac{1+\xi^2}{1-\xi}\right]_+\ln\frac{Q}{Q_t}+
\\+1-\xi-(1+\xi^2)\Bigg[\frac{\theta(2\xi-1)}{1-\xi}
\ln\frac{1-\xi}{\xi}\Bigg]_+
-(12\xi^3-10\xi^2+1)\Bigg[\frac{\theta(2\xi-1)}{2(1-\xi)}
\Bigg]_+\,,
\end{multline}
\begin{multline}
\label{eq:fg} \mathcal{F}_{g,2}\left(\xi,\frac{Q}{Q_t}\right)=
-2[\xi^2+(1-\xi)^2]\ln\frac{Q}{Q_t}+2\xi(1-\xi)+
\\+\left\{(2\xi-1)(-6\xi^2+6\xi-1)
-[\xi^2+(1-\xi)^2]\ln\frac{1-\xi}{\xi}\right\}
\theta\left(2\xi-1\right),
\end{multline}
\begin{eqnarray}
\mathcal{F}_{q,L}&=&
2\xi\left(2\xi-1\right)\theta\left(2\xi-1\right),
\end{eqnarray}
\begin{eqnarray}
\mathcal{F}_{g,L}&=&4\xi(1-\xi)(2\xi-1)\theta\left(2\xi-1\right).
\end{eqnarray}

The $(2n_f+1)\times 1$ matrices $\textbf{\emph{C}}_{0}$ and
$\textbf{\emph{C}}_{1}$ are defined such that their transposes are
given by:
\begin{equation}
\textbf{\emph{C}}_{0}^T(z)= \left(
  \begin{array}{c}
    e_u^2\,\delta(1-z) \\
    e_u^2\,\delta(1-z) \\
    \vdots \\
    0 \\
  \end{array}
\right),
\end{equation}
and:
\begin{equation}
\textbf{\emph{C}}_{1}^T(z)= \left(
  \begin{array}{c}
    C_F\,e_u^2 \left\{\mathcal{F}_{q,2}(z,1)-\frac{y^2}{1+(1-y)^2}\mathcal{F}_{q,L}(z,1)\right\} \\
    C_F\,e_u^2 \left\{\mathcal{F}_{q,2}(z,1)-\frac{y^2}{1+(1-y)^2}\mathcal{F}_{q,L}(z,1)\right\} \\
    \vdots \\
    T_f\left\{\mathcal{F}_{g,2}(z,1)-\frac{y^2}{1+(1-y)^2}\mathcal{F}_{g,L}(z,1)\right\} \\
  \end{array}
\right).
\end{equation}

\section{The radiator}
\label{sec:rad}

The radiator is given by:
\begin{equation}
R(Q/Q_t)= L g_1(\alpha_sL)+g_2(\alpha_sL)\,,
\end{equation}
where $L=\ln(Q/Q_t)$ and in the $\overline{\textrm{MS}}$ scheme:
\begin{equation}
g_1(\alpha_sL)=-\frac{C_F}{\pi\beta_0}
\left[1+\frac{\ln(1-2\lambda)}{2\lambda}\right],
\end{equation}
\begin{multline}
g_2(\alpha_sL)=\frac{3C_F}{4\pi\beta_0}\ln(1-2\lambda)
+\frac{C_FK}{4\pi^2\beta_0^2}\left[\ln(1-2\lambda)
+\frac{2\lambda}{1-2\lambda}\right]-\\
-\frac{C_F\beta_1}{2\pi\beta_0^3}
\left[\frac{2\lambda+\ln(1-2\lambda)}{1-2\lambda}
+\frac{1}{2}\ln^2(1-2\lambda)\right],
\end{multline}
with $\lambda=\alpha_s(Q^2)\beta_0L$ and:
\begin{equation}
K=C_A\left(\frac{67}{18}-\frac{\pi^2}{6}\right)-\frac{5}{9}n_f\,
,\qquad \beta_0=\frac{11C_A-2n_f}{12\pi}\, ,\qquad\beta_1=\frac{17
C_A^2-5C_An_f-3C_Fn_f}{24\pi^2}\,.
\end{equation}
In obtaining this, we used Eq.\eqref{eq:rad10} and the 2-loop QCD
$\beta$ function to replace the scale of $\alpha_s$ with $Q^2$  and
moved from the CMW \cite{CMW} scheme to the $\overline{\textrm{MS}}$
scheme (see e.g. Ref.~\cite{brd}). The derivative of the radiator
with respect to $\ln b$ at $\bar{b}\,Q=Q/Q_t$ is given by:
\begin{equation}
R'(Q/Q_t)=\frac{2C_F}{\pi\beta_0}\frac{\lambda}{1-2\lambda}\,.
\end{equation}

The expansion of the resummed result (Eq.\eqref{eq:res1}) to
$\mathcal{O}(\bar{\alpha}_s)$ and $\mathcal{O}(\bar{\alpha}_s^2)$,
which is needed in Eq. \eqref{eq:matching}, yields:
\begin{equation}
\sigma_r^{(1)}/\sigma_0=G_{11}L+G_{12}L^2
 -2\frac{\textbf{\emph{C}}_0\otimes\textbf{\emph{P}}\otimes\textbf{\emph{q}}(x,Q^2)}{q(x,Q^2)}L
 +\frac{\textbf{\emph{C}}_1\otimes\textbf{\emph{q}}(x,Q^2)}{q(x,Q^2)}\, ,
\end{equation}
\begin{multline}
\sigma_r^{(2)}/\sigma_0=G_{22}L^2+G_{23}L^3+\frac{1}{2}G_{12}^2L^4
-2\left((G_{11}+2\pi\beta_0)L^2+G_{12}L^3\right)
\frac{\textbf{\emph{C}}_0\otimes\textbf{\emph{P}}
\otimes\textbf{\emph{q}}(x,Q^2)}{q(x,Q^2)}
\\+\frac{\textbf{\emph{C}}_1\otimes\textbf{\emph{q}}
(x,Q^2)}{q(x,Q^2)} \left(G_{11}L+G_{12}L^2\right) +2
\frac{\textbf{\emph{C}}_0\otimes\textbf{\emph{P}}
\otimes\textbf{\emph{P}}\otimes\textbf{\emph{q}}(x,Q^2)}{q(x,Q^2)}
L^2\, ,
\end{multline}
where $\textbf{\emph{P}}$ is the matrix of leading order splitting
functions and the coefficients $G_{mn}$ are given in table
\ref{tab:coefs}. One can clearly see that the expansion of the
resummed result to $\mathcal{O}(\alpha_s)$ reproduces the leading
order result given by Eq.~\eqref{eq:LO}.

\renewcommand{\arraystretch}{1.2} \TABLE[ht]{
\begin{tabular}{|c|l|}
 \hline
  $\,G_{12}\,$ & $\,-2\,C_F$ \\ \hline
  $\,G_{11}\,$ & $\,3\,C_F$ \\ \hline
  $\,G_{23}\,$ & $\,-6\,C_F^2-\frac{16}{3}\pi\, C_F\beta_0$ \\ \hline
  $\,G_{22}\,$ & $\,-\frac{\pi^2}{3}C_FC_A+\frac{9}{2}\,C_F^2-2\,C_FK+6\,\pi\,C_F\beta_0$\qquad\,\\
  \hline
\end{tabular}
\caption{\label{tab:coefs}The coefficient $G_{nm}$ that enter the
fixed-order expansion of the resummed result.}} 

\end{document}